# Audio Steganography: LSB Technique Using a Pyramid Structure and Range of Bytes

Satish Bhalshankar[1], Avinash K. Gulve[2]

Research Student, Dept. of Computer Science and Engineering, Government College of Engineering, Aurangabad[1]
Associate Professor, Dept. of Computer Science and Engineering, Government College of Engineering, Aurangabad[2]

## Abstract

*The demand for keeping the information secure and confidential simultaneously has been progressively increasing. Among various techniques- Audio Steganography, a technique of embedding information transparently in a digital media thereby restricting the access to such information has been prominently developed. Imperceptibility, robustness, and payload or hiding capacity are the main character for it. In earlier, LSB techniques increased payload capacity would hamper robustness as well as imperceptibility of the cover media and vice versa.*

*The proposed technique overcomes the problem. It provides relatively good improvement in the payload capacity by dividing the bytes of cover media into ranges to hide the bits of secret message appropriately. As well as due to the use of ranges of bytes the robustness of cover media has maintained and imperceptibility preserved by using a pyramid structure.*

## 1. Introduction

Steganography is the adroit skill to cloak data in a cover media such as text, audio, image, video, etc. The term steganography derived from Greek which means, "Covered Writing". Steganography is the one of the major techniques in the area of information hiding. There are many stories about Steganography. For example ancient Greece used methods for hiding messages such as hiding it in the belly of a hare (a kind of rabbits), using invisible ink and pigeons. Another ingenious method was to shave the head of a messenger and tattoo a message or image on the messenger head. After allowing his hair to grow, the message would be undetected until the head was shaved again Steganography provides techniques for masking the existence of a secondary message in the presence of a primitive message. The primitive message is accredited to as the carrier signal or carrier message, the carrier signal can be text, audio, image, video, etc., the secondary message is assigned to as the payload signal or payload message. The message is being hidden in such a way that the presence of a secondary message is unrecognized to the onlooker and the carrier signal is modified in an imperceptible manner as shown in Figure 1.

Generally, Cryptography involves the encryption of the message. It makes no attempt to hide the encrypted message.

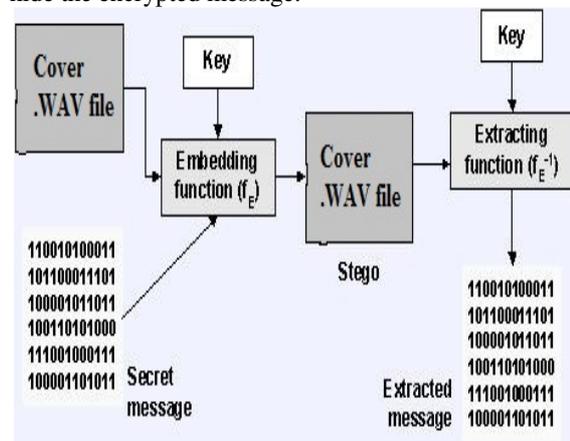

**Figure 1: Audio Steganography System (Stego-system)**

In Steganography, the original message is not altered but the very existence is hidden from the observer by embedding the message in the selected medium.
Audio steganography:
Cover signal + Target data = Stego signal (Transmitted)

There are distinct steganographic methods for masking the furtive message. The principal requirement for a steganographic method is imperceptibility which means that the furtive messages should not be discernible to the human by vision or audio. There are two more requirements, one is to maximize the hiding capacity, and the other is protection. In Steganography, one technique where using audio files as stego-object. In a computer-based audio steganography system, digital sound is used for masking the furtive message.

By slightly varying the binary sequence of a sound file, the secret message is embedded into the audio data file. In the last few years, various algorithms have been developed for the embedding and extraction of a message in audio signals. All of the developed algorithms take advantage of the perceptual properties of the human auditory system (HAS) in order to add a message into a host signal in a perceptually transparent manner. Hiding extra information into audio signals is a little bit interesting but suspicious, as Human Auditory System (HAS) is more sensitive than Human Visual System (HVS) [1].

The masking of the confidential data into the secret medium should not make any loathsome changes to the secret medium so that the authenticity of the file should not disturbed. The audio steganography view is to ingrain valuable confidential data into an audio file in such a way that human auditory system (HAS) cannot to detect the change which occurred due to ingraining of the data into the audio file. In the audio steganography, Least Significant Bit (LSB), Spread spectrum, and Echo hiding approaches along with other current applications that have been developed in recent years. The properties of audio steganography [2] being exploited in different steganography applications are

    a. Confidentiality
    b. Imperceptibility
    c. High capacity
    d. Difficult Detectability
    e. Accurateness
    f. Survivability
    g. Visibility

Audio steganography is found to be durable and strong avenue auditory system is much wiser than the human visual system. The idea is to ingrain the secret data into an audio file such that there is the imperceptible difference between the original audio file and embedded file. While embedding the furtive data the format has to be keep in mind so that header part of the wave file (first 44 byte) [3] should be untouched because in case the header gets corrupted, the audio file will also corrupt as shown in Figure 2. The second consideration that should be made is not to embed data into the silent zone as that might cause undesirable change to the audio file.

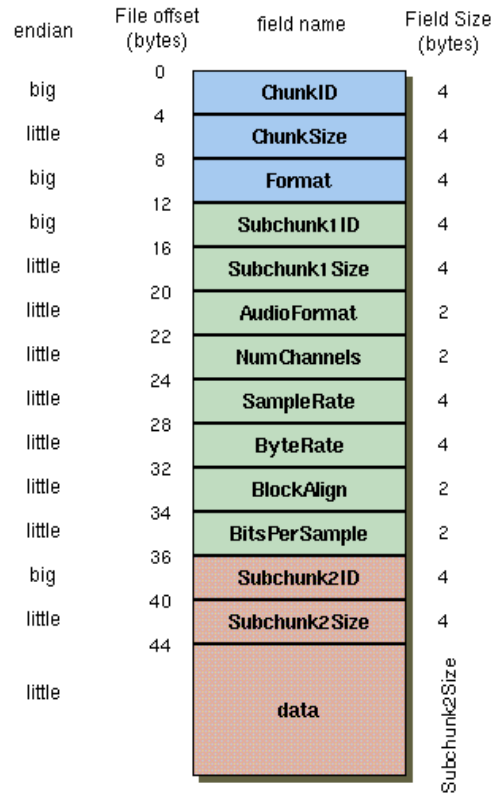

**Figure 2: Wav File Format**

## 2. Aged Techniques

In audio Steganography, a furtive message is embedded into a digitized audio signal which results in slight altering of the binary sequence of the consequent cover audio file. There are numerous procedures are available for audio. Generally following types are useful in Audio Steganography:
1. Echo Hiding
2. Phase Coding
3. Spread Spectrum
4. LSB Coding

In echo hiding, furtive information is embedded in a .wav audio file by producing an echo into the discrete signal. Three parameters of the echo are assorted: amplitude, decay rate, and offset from the original signal. All three parameters are set below the human hearing threshold so the echo is not determined. In short, one echo was produced from the original signal. Then one bit of information could be encoded. That's why; the original signal is broken down into blocks before the encoding process begins. Once the encoding process is completed, the blocks are concatenated back together to create the final signal. To extract the furtive message from the stego-signal, the receiver must be able to break up the signal into the same block sequence used during the encoding process.

In Phase coding, relies on the fact that the phase components of audio signals are not as detectable to the human ear as noise is. It is based on replacing selected phase components from the original audio signal spectrum with hidden data. However, to guarantee inaudibility, phase components modification should be kept tiny.

The basic spread spectrum method attempts to spread secret information across the audio signal's frequency spectrum as much as possible. This is comparable to a system using an implementation of the LSB coding that randomly spreads the message bits over the entire sound file. This method spreads the secret message over the sound file's frequency spectrum, using a code that is independent of the actual signal. As a result, the final signal occupies a bandwidth in surplus of actual required for transmission.

Finally which is considered for research work, i.e. Least Significant Bit (LSB) coding is another way to embed information in a digital audio file. One of the most primitive techniques considered in the information hiding of digital audio as well as other media types is LSB Embedding. In this technique, LSB of a binary sequence of each sample of the digitized audio file is replaced with the binary equivalent of secret data. That's usually an effective technique in cases where the LSB substitution doesn't cause significant quality deprivation.

By substituting the least significant bit of each sampling point with bits of a secret message, LSB coding permits embedding of secret data in a better quantity. In some implementations of LSB embedding, however, the 1 to 4 least significant bits of a sample are replaced with 1 to 4 message bits. The large quantity of secret data gets hiding but also increases the amount of resulting noise in the audio file as well. Thus, it concerns to choose the signal content before deciding on the LSB operation to use.

To extract a furtive message from an LSB encoded sound file, the recipient needs access to the sequence of sample indices used in the embedding process. Normally, the length of the furtive message to be encoded is smaller than the total number of samples in an audio file. One must decide then on how to choose the subset of samples that will contain the furtive message and communicate that decision to the recipient. In LSB technique, skip the beginning of the sound file after that perform LSB coding until the message has been completely embedded, leaving the lingering samples untouched. This generates a security problem; however the first part of the sound file will have different statistical properties than the second part of the sound file that was not modified. An answer to this problem is to protect the furtive message with random bits so that the length of the message is equal to the total number of samples. Still the embedding process ends up changing far more samples than the transmission of the secret required. This increases the probability that a would-be attacker will suspect furtive communication.

For example, to hide the letter "A" (ASCII code 65, which is 01000001) inside eight bytes of a cover, set the LSB of each byte by selecting one bit of the text data at a time and correcting the LSB of the envelope data bytes accordingly as follows.

Original Audio Bytes Text data to hide Text data Embedded Audio Bytes

| Original Audio Bytes | Text data to hide | Embedded Audio Bytes |
|---|---|---|
| 1001001**0** | 0 | 1001001**0** |
| 0101001**1** | 1 | 0101001**1** |
| 1001101**1** | 0 | 1001101**0** |
| 1101001**1** | 0 | 1101001**0** |
| 1000101**0** | 0 | 1000101**0** |
| 0000001**1** | 1 | 0000001**0** |
| 0111001**0** | 0 | 0111001**0** |
| 0010101**0** | 0 | 0010101**1** |

*P. G. Mamatha, T. Ravi Kumar Naidu, T.V.S. Gowtham Prasad [4]* had implemented this technique. Wherein, LSB coding gave high bit rate was easy for implementation and easy to detect. Author had recommended the LSB technique with XORing for improvement in security. This method suppose to perform XOR operation on the LSBs and depending on the result of XOR operation and the message to be embed, the LSB of the sample might be modified or remains same.

The LSB bits were flipped only when current bit with next bit using XOR operation between them. Author had analyzed their technique by performing MSE and PSNR tests on sample wav files.

One of the observations was that the values of PSNR test had decreased as payload capacity increased. The range of PSNR test changed from 36.70 to 28.32 as size of secret data increased.

*Neha Gupta and Nidhi Sharma [5]* had proposed technique using DWT and LSB. In this technique, for embedding the image in audio author considered the concept of least significant bit by using DWT.

- **Hiding Process:**

  Author converted cover file (i.e. Audio file) into byte format and secret file (i.e. Image file) into bits format subsequently. After that author applied the DWT (Discrete wavelet transforms) on audio files for taking the higher frequency and generated a random key. After that author took 8x8 blocks for each 16 bits data and stored the image bits into the last 3 bits of the audio file.

- **Extraction Process:**

  Steps in the extraction process are opposite to that of embedding process.

  Analysis of this technique is that the size of secret data was too short. Therefore the payload capacity which is one of main characters of Steganography was not achieved. As well as the values of PSNR test for storing the same secret data into different audio files fluctuated.

  *Padmashree G and Venugopala P S [11]* also did their research work on LSB Audio Steganography. They had used 4th and 5th Layers of Bytes of Cover Media File.

  In this technique, on dispatcher side, the text file which had embedded into an audio file was encrypted using a public key cryptographic algorithm, RSA. The cipher text obtained was then embedded in the 4th and 5th LSB bit using LSB algorithm. The stego audio file contains the covert message embedded into it.

  On the recipient side, the embedded audio file was selected to extract the covert message. The covert message was decrypted using RSA decryption method.

  One of the observations is that the authors had performed MSE, PSNR, and SNR tests on various audio files. The range of PSNR test was from 10 to 17. And the size of the covert media file had not mentioned, therefore the payload capacity is unpredictable.

## 3. Proposed Algorithm

The improved method for audio steganography is proposed in this paper. The proposed method improves the hiding capacity with fewer distortions in signals of original sound file. The combination of a Pyramid Structure and Range of Bytes gave good result, because the furtive data is hiding in cover audio file randomly with variation. The slight modification in LSBs is suggested, to preserve the imperceptibility.

Figure 3 depicts the workflow of Embedding Process for the proposed algorithm.

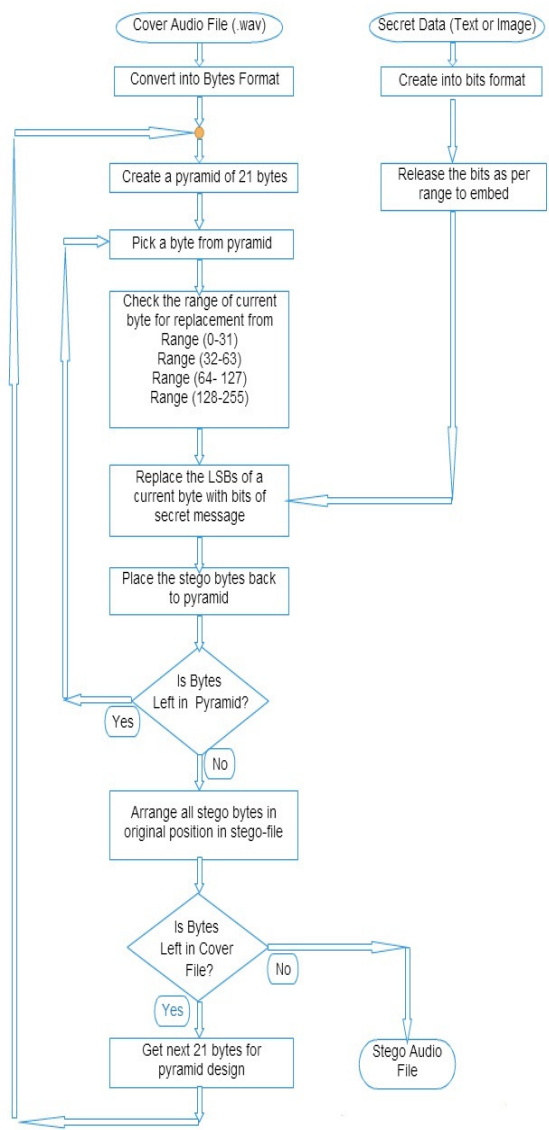

**Figure 3: Work flow diagram for Embedding Process of Proposed Algorithm**

Proposed Algorithm uses range of bytes of cover audio file to hide bits of secret information. But before replacing the LSBs of the selected bytes to ensure the arbitrariness, a design which embeds furtive data bits in LSBs or higher layers based on ranges of bytes [10] is proposed.

Figure 4 shows how to pick the byte for storing purpose.

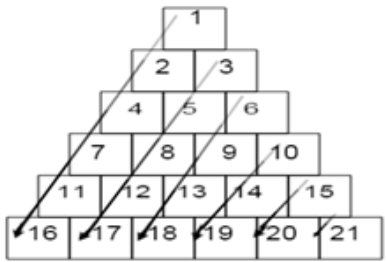

**Figure 4: Pyramid structure of bytes**

The sequence of picking the bytes from different ranges for embedding plays an important role in this algorithm. For e.g., Range is (32-63) with byte value is 63 of cover file, whereas form secret file the bit pattern is 011.
Then
    0011 1**111** = **63**  (Before embedding)
         **011** = **03**  (No. of Bits for Replacement)

    0011 1**011** = **61** (After embedding)

As seen here, even the current byte is not having much variation after replacing the three LSBs. Now as and when Range (i.e. in between 0 to 255) is increasing then the replacement of maximum bits (From $2^{nd}$ to $4^{th}$ layer) is also possible. So ultimately the payload or hiding capacity will be enhanced.

This method has increased level of security in subsequent LSB modifications with Pyramid Structure.

Following is the step by step process for embedding and extraction.

- **Embedding Process:**
1. Take Audio (.wav file) as Cover Media.
2. Convert the Audio File into Bytes Format.
3. Take Secret File (Text or Image Data).
4. Convert the Secret File into Bits Pattern.
5. Now Skip the First 44 Bytes of Audio File and Create a Pyramid of next 21 bytes as shown in Figure 3.
6. Pick a byte from pyramid.
7. After picking a byte, check the range of that byte from the Ranges predefined.
8. Then replace the respective LSBs of the current byte using the bits from secret file.
9. Do the **Step 6 to Step 8** until all the bytes of the pyramid will get visited.
10. After that arrange the bytes from pyramid into their original position in Cover audio file.
11. Then check whether bytes are remaining in Cover file. If yes, get next 21 byte for pyramid and repeat the process else stop.
12. Then attach the first 44 bytes with modified Bytes and Convert the Bytes Format into Stego Audio File.

In the above process, Cover Audio File should have sufficient samples to embed the secret file otherwise algorithm must give an error message.

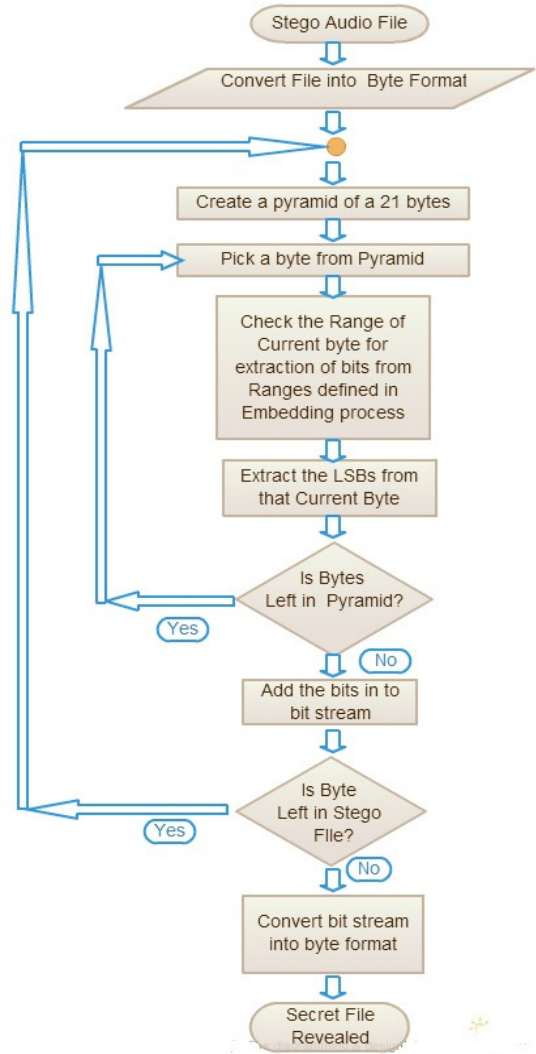

**Figure 5: Work flow diagram for Extraction Process of Proposed Algorithm**

- **Extraction Process:**
1. Take Stego Audio (.wav file).
2. Convert the Stego Audio File into Bytes Format.
3. Now Skip the First 44 Bytes of Audio File and Create a Pyramid of next 21 bytes as shown in Figure 5.
4. Pick a byte from pyramid.
5. After picking a byte, check the range of that byte from the Ranges predefined.
6. Then extract the respective LSBs of the current byte using the bits from Stego File.

7. Do the **Step 4 to Step 6** until all the bytes of the pyramid will get visited.
8. After that, add the extracted bits into bit pattern or stream.
9. Then check whether bytes are remaining in Stego file. If yes, get next 21 byte for pyramid and repeat the process else stop.
10. Then Convert the bit pattern into Bytes format.

Finally Convert the Bytes Format into Secret Message File.

## 4. Performance Analysis

This Steganography Technique is implemented in Visual C# 2010. The efficiency of steganography algorithm can be gauged subject to fulfill of some basic requirements. The requirements are nonappearance of secret data, hiding capacity, robustness against malicious attacks and independent of file format. In this algorithm, wav audio file format [3] has used as cover media. The Peak Signal Noise Ratio (PSNR), Mean Square Error (MSE) and Payload capacity of wav audio format is calculated and compared using different music genre. Finally, the histograms which are designed with the help of Zero Crossing Rate (ZCR) technique of cover audio and stego audio have been compared.

The above tests carried out for the above algorithm using MATLAB R2009 with various different wav audio files.

**MSE (Mean Square Error)**
This is the first test used for performance analysis where 255 is the highest value of audio intensity and MSE (Mean Square Error) [6] is the average value of the total square of Absolute Error between cover file and stego file. MSE can be counted with the formula bellow:

$$\text{MSE}(x,y) = \frac{1}{N}\sum_{i=1}^{N}(xi - yi)2 \qquad 1$$

**PSNR (Peak Signal Noise Ratio)**
This Steganography research will test the level of quality stego file after the message has embedded in original file. The 8-bit and 16-bit wav format file has tested using the Peak Signal to Noise Ratio (PSNR) [6] [9] formula which will be counted in decibel (dB). The value of PSNR is good if it is above of 20 dB with formula.

$$\text{PSNR} = 10\log_{10}\left(\frac{255}{\text{MSE}}\right) \qquad 2$$

**ZCR (Zero Crossing Rate)**

Zero-Crossing Rate [7] [8] is a measure of the number of times in a given time interval that the amplitude of the speech signals passes during a value of zero. Because of its random nature, the zero-crossing rate for unvoiced speech is greater than that of voiced speech. The zero-crossing rate is an important parameter for voiced/unvoiced classification and for endpoint detection. Detecting a speech utterance begins and ends is a basic problem in speech processing. This is often referred to as endpoint detection. End-point finding is complicated if the speech is uttered in a noisy environment.

It indicates the frequency of signal amplitude sign changes. To some extent, it indicates the average signal frequency as:

$$\text{ZCR} = \frac{\sum_{n=1}^{N}|sgn\,x(n) - sgn\,x(n-1)|}{2N} \qquad 3$$

For experiment, initially 8 Bit and 16 Bit uncompressed Wav files used as cover media and text files as a secret.

MSE serves as an important parameter in gauging the performance of the steganographic system. Suppose that x = {xi | i = 1, 2. . . N} and y = {yi | i = 1, 2. . . N} are two finite-length, discrete signals, for e.g., images and audio signals. Then MSE calculation between the signals is given by equation (1). The following table 1 gave experimental result of MSE values

**Table 1: MSE Calculation**

| Music Genre | File name | MSE (8 Bit) | MSE (16 Bit) |
|---|---|---|---|
| Bass | Bass1 | 0.000628 | 0.00044 |
| | Bass2 | 0.000518 | 0.000453 |
| | Bass3 | 0.000721 | 0.000245 |
| Drum | Drum1 | 0.000798 | 0.00054 |
| | Drum2 | 0.000643 | 0.000879 |
| | Drum3 | 0.000727 | 0.000529 |
| Music Genre | File name | MSE (8 Bit) | MSE (16 Bit) |
| Dance | Dance1 | 0.000991 | 0.00063 |
| | Dance2 | 0.000637 | 0.000663 |
| | Dance3 | 0.000699 | 0.000763 |
| HipHop | HipHop1 | 0.000827 | 0.000633 |
| | HipHop2 | 0.000774 | 0.00066 |
| | HipHop3 | 0.000653 | 0.000451 |
| Rock | Rock1 | 0.000825 | 0.000639 |
| | Rock2 | 0.000693 | 0.000698 |
| | Rock3 | 0.000742 | 0.000519 |
| Voice | Voice1 | 0.000755 | 0.000598 |
| | Voice2 | 0.000770 | 0.000542 |
| | Voice3 | 0.000960 | 0.000483 |
| Animal | Animal1 | 0.000819 | 0.000706 |
| | Animal2 | 0.000825 | 0.000528 |

| | Animal3 | 0.000729 | 0.00066 |

Here, the MSE values for files of both types (8 Bit and 16 Bit) are given which are much less as expected.

The MSE values are relatively good as compared to the technique of *P.G. Mamatha [4]* and et al.

The following charts make it more clearly about MSE. The MSE values are ranging from 0.0005 to 0.001.

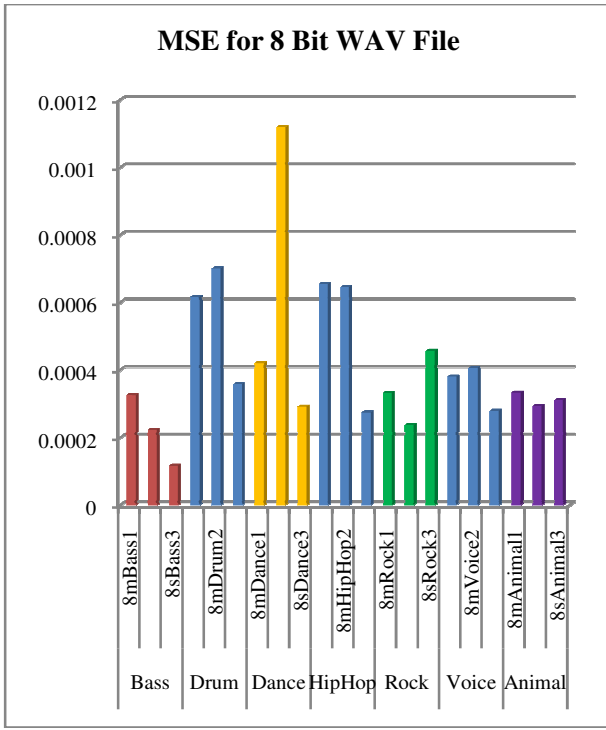

**Figure 6: MSE Values with respect to secret file hiding
(8 Bit Wav Files)**

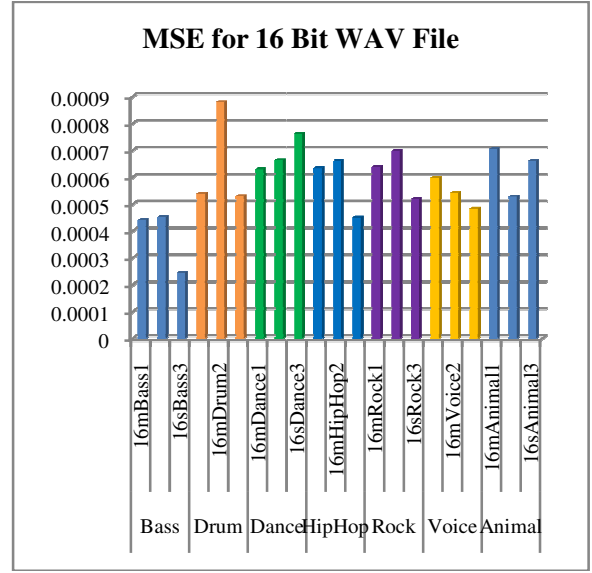

**Figure 7: MSE Values with respect to secret file hiding
(16 Bit Wav Files)**

The second test which is next step of MSE i.e. PSNR is also having major importance by comparing the original file with stego file. In the experiment, the PSNR range found between 54 to 60 dB by using this technique for both types (i.e. 8 Bit and 16 Bit).

The following graphs in figure 9 and 10, showing the analysis about the PSNR values for 8 bit and 16 bit audio files.

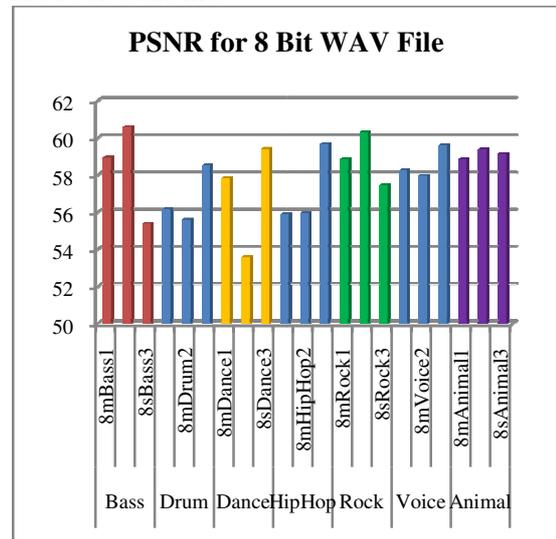

**Figure 8: PSNR Values with respect to secret file hiding
(8 Bit Wav Files)**

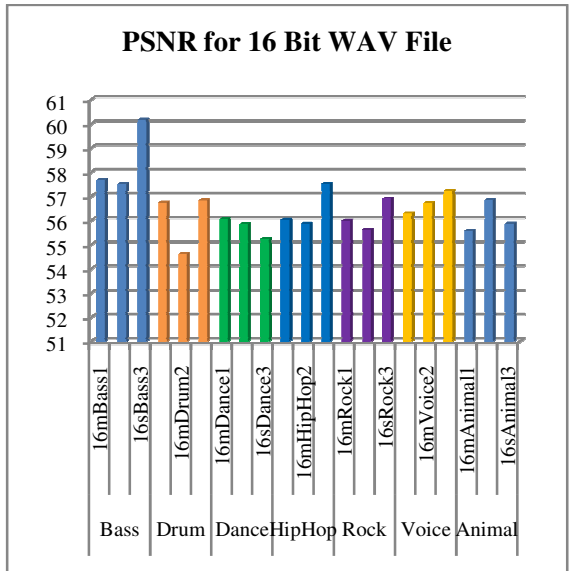

**Figure 9: PSNR Values with respect to secret file hiding
(16 Bit Wav Files)**

The observation of individual 8 bit and 16 bit original audio files with stego files gave a good result with better decibel values as shown in figure 8 and 9.

The one more observation is that the proposed technique gave averagely a good result with all types of music genre used in experiment with respect to PNSR. As well as it is another observation is that the highest PNSR 60.59db for the 8 bit Bass2 file which is the top most PNSR among files and music genre.

But if PSNR values of 8-Bit compared with 16-Bit then the analysis and observation is as follows:

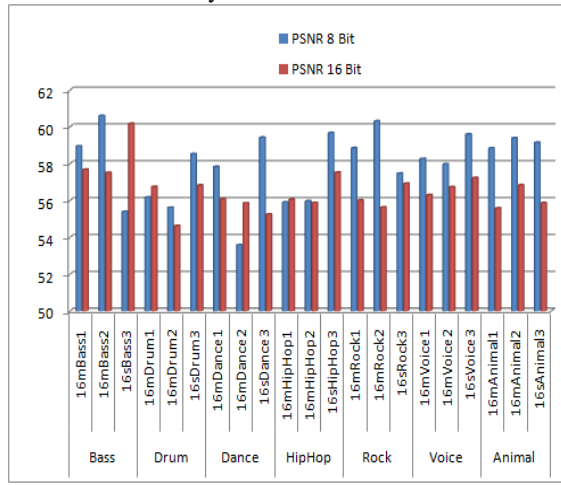

**Figure 10: PSNR Comparison between 8 Bit and 16 Bit WAV Files**

In Figure 10, the average PSNR of 8 bit audio files is 58.05 while for 16 bit audio files is 56.53. Due to these values, the main observation is that this technique works with 8-bit files somehow better than 16-bit files of all types music genre used in the experiment.

Overall, due to the good results of PSNR values the imperceptibility maintained which is one the major factors of steganography.

Next is Payload Capacity which has its own significance in Steganography. Because hiding the bits of furtive message in cover sound or media file without disturbing quality of sound signals. And this is the big challenge. In the proposed and implemented algorithm the payload capacity achieved as follows.

The following table 2 gives information about PSNR as well as the payload capacity of 8 bit files while table 3 gives of 16 bit wav files. In both table, third column shows size of cover media file while fourth shows secret message file size. As well as fifth column shows percentage ratio of payload capacity with respect to the proposed technique.

The important observation about the payload capacity is that the payload capacity also better. Payload capacity ranged from 17 % to 21% for 8 bit audio files while for 16 bit audio files it is 19 % to 24% maintained.

Here averagely, the payload capacity of 8 bit audio files is 18.64% while for 16 bit audio files it is 20.71%. So, 16 bit files having better result than 8 bit files for payload capacity parameter where as the highest payload capacity of this algorithm is 24.03%

**Table 2: Payload Capacity and PSNR of 8-Bit Wav Files**

| Music Genre | File name | WAV Size (Byte) | Message Size (Byte) | Payload Capacity Percentage (%) | PSNR |
|---|---|---|---|---|---|
| Bass | Bass1 | 66156 | 12030 | 18.18 | 58.944 |
| Bass | Bass2 | 86060 | 16122 | 18.73 | 60.596 |
| Bass | Bass3 | 18964 | 3230 | 17.03 | 55.386 |
| Drum | Drum1 | 27042 | 5094 | 18.83 | 56.164 |
| Drum | Drum2 | 16957 | 3162 | 18.64 | 55.607 |
| Drum | Drum3 | 49030 | 9280 | 18.92 | 58.529 |
| Dance | Dance1 | 65942 | 12380 | 18.77 | 59.78 |
| Dance | Dance2 | 30142 | 5633 | 18.68 | 57.833 |
| Dance | Dance3 | 9993 | 1856 | 18.57 | 53.578 |
| Music Genre | File name | WAV Size (Byte) | Message Size (Byte) | Payload Capacity Percentage (%) | PSNR |

| Music Genre | File name | WAV Size (Byte) | Message Size (Byte) | Payload Capacity Percentage (%) | PSNR |
|---|---|---|---|---|---|
| HipHop | HipHop1 | 57194 | 10725 | 18.75 | 59.415 |
| | HipHop2 | 18924 | 3520 | 18.60 | 55.902 |
| | HipHop3 | 18924 | 3523 | 18.61 | 55.970 |
| Rock | Rock1 | 61190 | 11515 | 18.81 | 58.849 |
| | Rock2 | 69502 | 13060 | 18.79 | 60.317 |
| | Rock3 | 30316 | 5682 | 18.74 | 57.463 |
| Voice | Voice1 | 37339 | 6972 | 20.22 | 58.256 |
| | Voice2 | 35602 | 6708 | 18.84 | 57.976 |
| | Voice3 | 56270 | 10554 | 18.75 | 59.599 |
| Animal | Animal1 | 41898 | 7839 | 18.71 | 58.842 |
| | Animal2 | 59160 | 11076 | 18.72 | 59.387 |
| | Animal3 | 46386 | 8681 | 18.71 | 59.142 |

**Table 3: Payload Capacity and PSNR of 16-Bit Wav Files**

| Music Genre | File name | WAV Size (Byte) | Mess-age Size (Byte) | Payload Capacity Percentage (%) | PSNR |
|---|---|---|---|---|---|
| Bass | Bass1 | 33060 | 6925 | 20.95 | 57.677 |
| | Bass2 | 124828 | 24845 | 19.90 | 57.508 |
| | Bass3 | 37887 | 7415 | 19.57 | 60.175 |
| Drum | Drum1 | 49706 | 10178 | 20.47 | 56.742 |
| | Drum2 | 31174 | 7494 | 24.03 | 54.626 |
| | Drum3 | 98016 | 19793 | 20.19 | 56.836 |
| Dance | Dance1 | 60240 | 12459 | 20.68 | 56.071 |
| | Dance2 | 19942 | 4020 | 20.15 | 55.846 |
| | Dance3 | 114344 | 27033 | 23.64 | 55.248 |
| Hip Hop | HipHop1 | 37804 | 7636 | 20.19 | 56.048 |
| | HipHop2 | 37804 | 8175 | 21.62 | 55.868 |
| | HipHop3 | 8784 | 1757 | 20.00 | 57.521 |
| Rock | Rock1 | 122336 | 24741 | 20.22 | 56.009 |
| | Rock2 | 138960 | 28002 | 20.15 | 55.629 |
| | Rock3 | 60588 | 12312 | 20.32 | 56.917 |
| Voice | Voice1 | 74634 | 15034 | 20.14 | 56.295 |
| | Voice2 | 71160 | 14677 | 20.62 | 56.724 |
| | Voice3 | 112496 | 22766 | 20.24 | 57.222 |
| Animal | Animal1 | 55840 | 11685 | 20.92 | 55.577 |
| | Animal2 | 78856 | 16097 | 20.41 | 56.838 |
| | Animal3 | 85304 | 17561 | 20.59 | 55.870 |

After comparing the PSNR values of proposed technique with P. G. Mamatha, T. Ravi Kumar Naidu, T.V.S. Gowtham's [4] technique which having PSNR range 28 to 36 db, and Neha Gupta and Nidhi Sharma's [5] technique which having PSNR max range up-to 37 db, as well as technique of Padmashree & Venugopala [11] having PSNR range at most 18 db, and the proposed algorithm having range 52 to 60 db means it shows the current technique is relatively good.

In the study, the next test performed drawing Histogram. The histogram helps to show the variation in signals. The histogram of original cover audio file and embedded stego files has generated with the help of Matlab. These histograms are really helpful to prove the proposed algorithm after implementation. Here some graphs are included as a proof that the technique is secure to transfer data.

An important point is to observe from these histograms is that, the proposed algorithm conserves the frequent shapes of the histograms. This feature of this technique makes it complicated to detect whether any data is hidden or not in the Stego Audio. Figure 11(a) & 11(b), Figure 12(a) & 12(b) and Figure 13(a) & 13 (b) showing the original cover file and embedded stego file. It can be observed that significant changes are not perceptible. Also the changes of pixel in graphs which represent the audio signal are not easily identifiable.

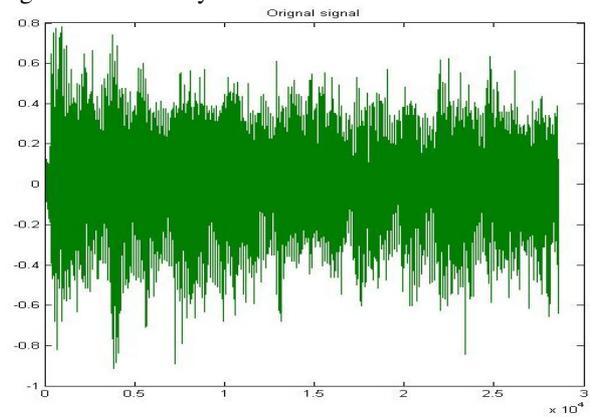

**Figure 11(a): Histogram of 8 Bit Dance3 Original Wav File**

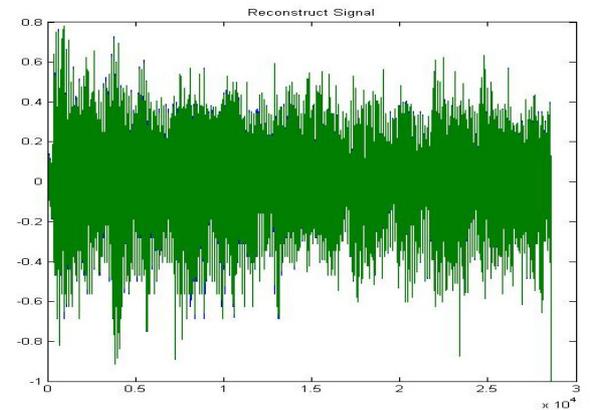

**Figure 11(b): Histogram of 8 Bit Dance3 Stego Wav File**

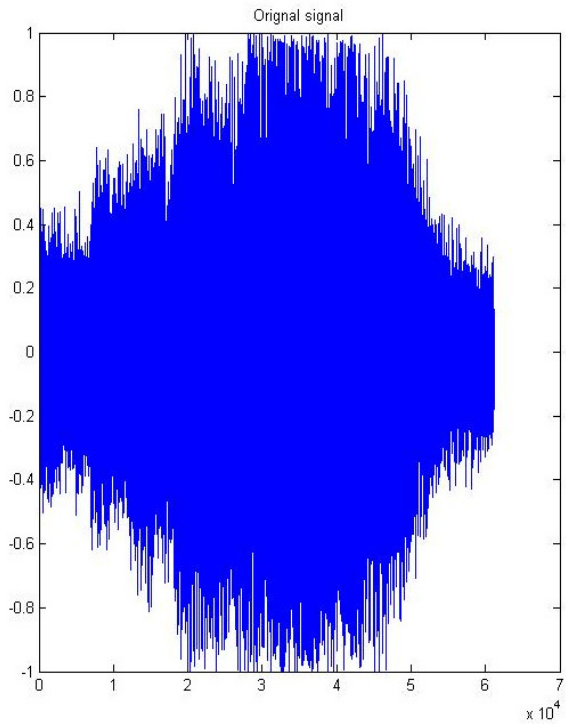

**Figure 12(a): Histogram of 16 Bit Rock1 Original Wav File**

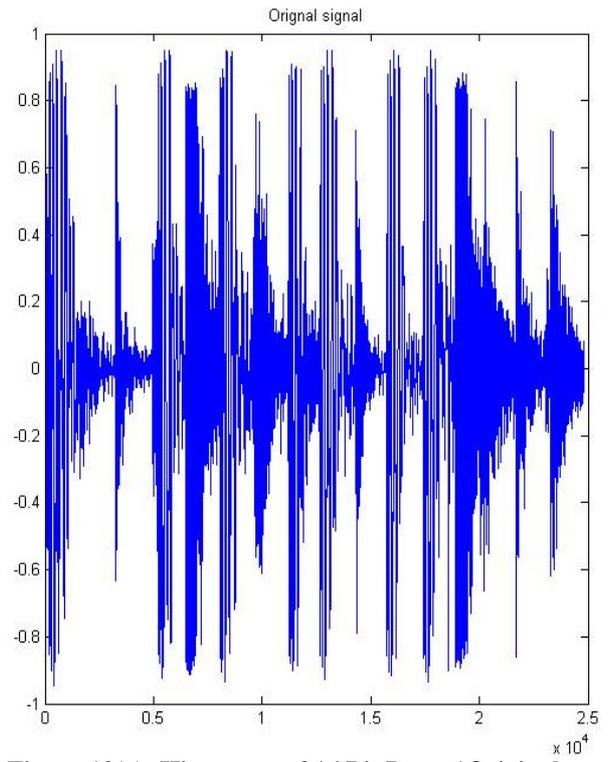

**Figure 13(a): Histogram of 16 Bit Drum1 Original Wav File**

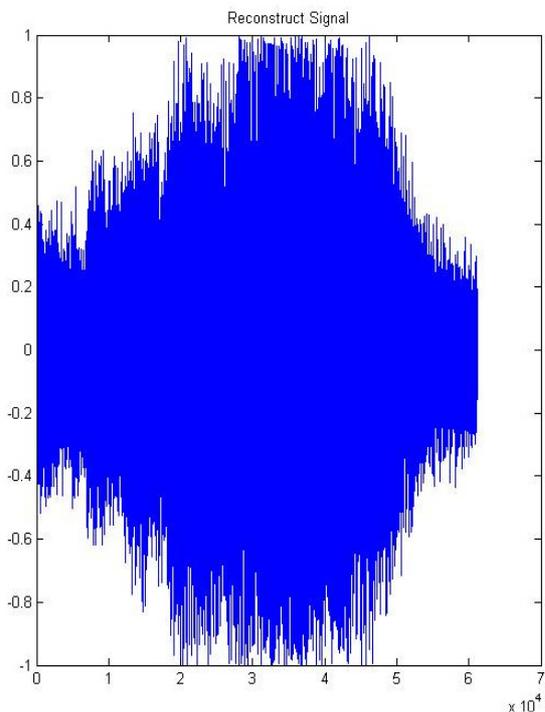

**Figure 12(b): Histogram of 16 Bit Rock1 Stego Wav File**

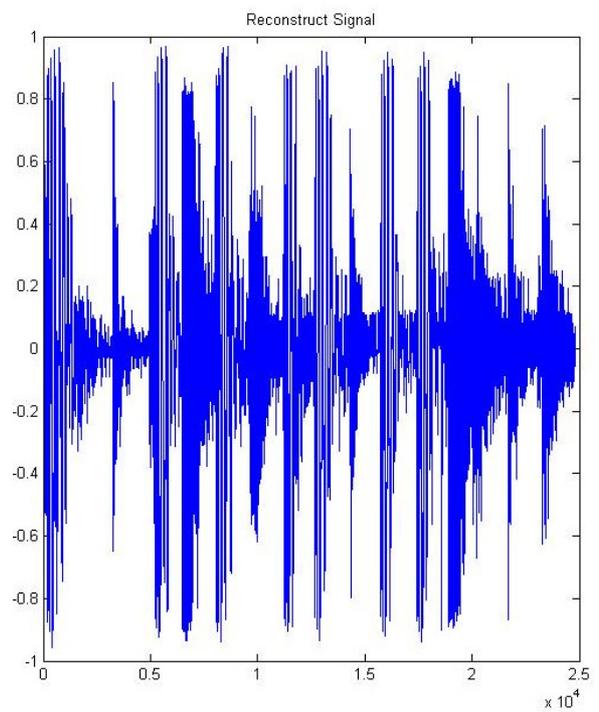

**Figure 13(b): Histogram of 16 Bit Drum1 Stego Wav File**

Zero-crossing rate (ZCR) is another basic auditory feature that can be calculated effortlessly. It is equal to the number of zero-crossing of the waveform within a given frame. ZCR has the following characteristics:

- ZCR of silent sounds and environmental noise are usually larger than voiced sounds, which has noticeable fundamental periods.
- It is hard to differentiate silent sounds from environmental noise by using ZCR alone since they have similar ZCR values.
- ZCR is often used in combination with energy (or volume) for end-point findings. In particular, ZCR is used for perceiving the start and end positions of silent sounds.

The following Graphs are examples of ZCR test.

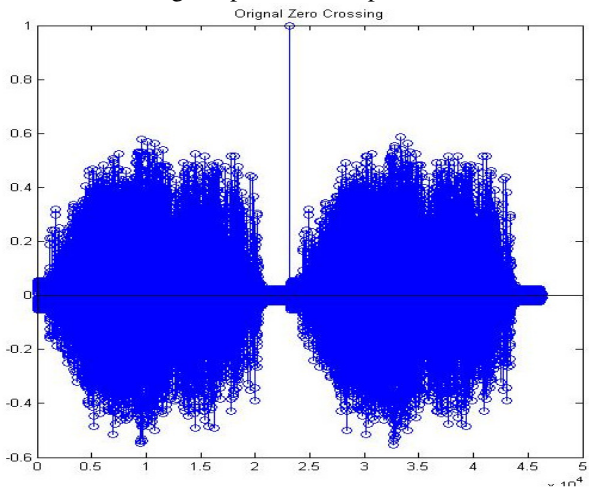

**Figure 14(a): ZCR Histogram of 8 Bit Animal3 Original Wav File**

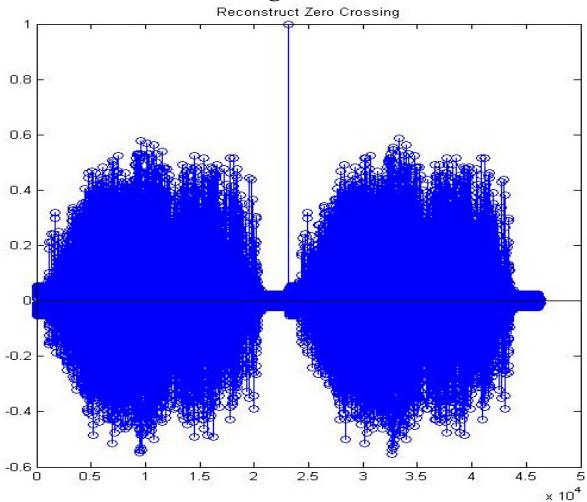

**Figure 14(b): ZCR Histogram of 8 Bit Animal3 Stego Wav File**

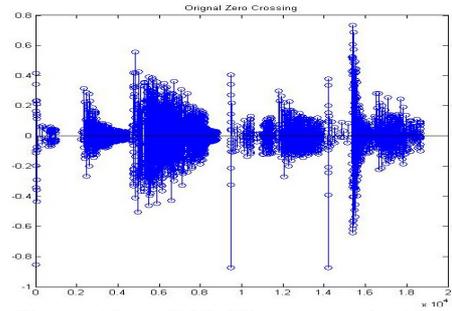

**Figure 15(a): ZCR Histogram of 16 Bit HipHop2 Original Wav File**

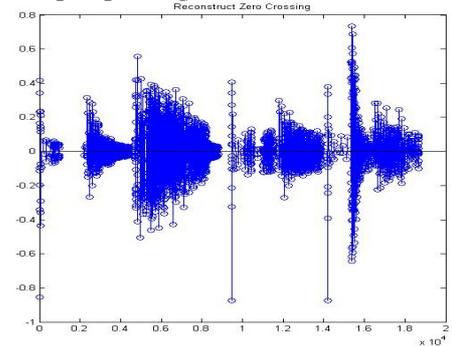

**Figure 15(b): ZCR Histogram of 16 Bit HipHop2 Stego Wav File**

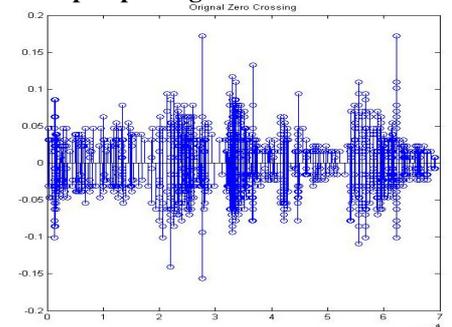

**Figure 16(a): ZCR Histogram of 8 Bit Rock2 Original Wav File**

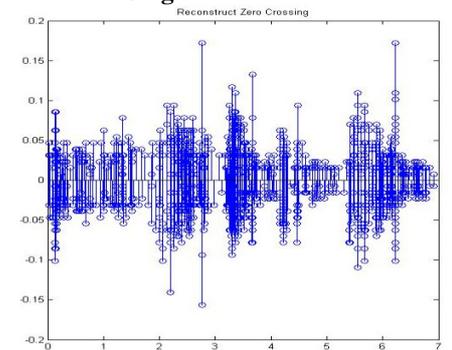

**Figure 16(b): ZCR Histogram of 8 Bit Rock2 Stego Wav File**

It is having an observation that the Figure 14(a) and 14(b) of 8 Bit Animal3 and Figure 15(a) and 15(b)

16 Bit HipHop2, also Figure 16(a) and 16(b) 8 bit Rock2 audio files are having the same structure. I.e. the observation is that Histograms generated using ZCR test for both files are equal.

## 5. Conclusion

The three pillars i.e. Robustness, Imperceptibility and Payload Capacity must be achieved for the success of good steganography.

The robustness of the proposed method is maintained due to use of a pyramid structure. The selection of bytes for embedding purpose (bytes of Cover Media File) is based on arithmetic progression which provides randomness.

The second pillar i.e. Imperceptibility, is also preserved by use of Range of Bytes to decide the number of secret data bits to be hidden. After picking a byte from pyramid, the value of byte is used for checking the range. Based on the values of bytes, number of bits used to hide secret data is estimated. Due to the use of Range of Bytes, replacement of LSBs varies from 0 to 4 bits which provides better imperceptivity.

The last pillar is the Payload Capacity. In the experiment, after skipping the first 44 bytes of .wav file (i.e. Cover Media File) the proposed method allowed the user to use all remaining bytes for embedding process. As well as use of range of bytes permit hiding of up-to 4 bits of secret message file in a byte of cover object. This provides better payload capacity which is up-to 24% of the cover media file size. This has been verified by carrying out experimentation on 42 .wav files.

## 6. Future Enhancement

Every algorithm needs improvement that's why this technique is also having future enhancement. Currently experiment is done on 8-bit and 16-bit wav files. Also another media files like mp3, mp4 can use as cover media. As well as improving the robustness is another challenge. Also, the incorporation of data compression and encryption techniques with the proposed algorithm may help to improve secrecy.

Satish Bhalshankar is currently pursuing his Master Degree in Computer Science and Engineering from Government College of Engineering, Aurangabad.
His area of interest is Information Security, Biometrics, Ethical Hacking.
Email: satish.bhalshankar@gmail.com

Avinash K Gulve is currently working as an Associcate Professor in the Department of Computer Science and Engineerin, Government College of Engineering, Aurangabad.
His area of interest is Information Security, Cryptography, Steganography, and Image Processing.
Email: akgulve@yahoo.com